\documentclass[prl,reprint,amsmath,amssymb,aps,preprintnumbers, longbibliography]{revtex4-1}

\usepackage[pdftex,     % sets up hyperref to use pdftex driver
            plainpages=false,   % allows page i and 1 to exist in the same document
            breaklinks=true,    % link texts can be broken at the end of line
            colorlinks=true
           ]{hyperref}

\usepackage{thumbpdf}
\usepackage{amsmath}
\usepackage{amssymb}
\usepackage{amsfonts}
\usepackage{array}
\usepackage{bbold}
\usepackage{cancel}
\usepackage{subfigure}
\usepackage{verbatim}
\usepackage{slashed}
\usepackage{graphicx}
\usepackage{url}
\usepackage{lipsum}
\usepackage{mathrsfs, bbm, braket, color}
\usepackage{soul}
\usepackage{psfrag}
\usepackage{amstext}
\usepackage{multirow,graphics}
\usepackage{enumerate}
\usepackage{color}
\usepackage{tikz}

\usepackage[utf8]{inputenc}

\setcitestyle{square,numbers}

\begin{document}

%NewCommands
\newcommand{\im}{{\rm i}}
\newcommand{\id}{{\mathbb 1}}
\newcommand{\nn}{{\nonumber}}
\newcommand{\p}[1]{\left({#1}\right)}
\newcommand{\comm}[1]{\left[{#1}\right]}
\newcommand{\acomm}[1]{\left\{{#1}\right\}}
\newcommand{\f}[2]{\frac{#1}{#2}}
\newcommand{\IM}{{\rm Im}\,}
\newcommand{\card}{\#}
\newcommand{\la}[1]{\label{#1}}
\newcommand{\eq}[1]{(\ref{#1})} 
\newcommand{\figref}[1]{Fig. \ref{#1}}
\newcommand{\abs}[1]{\left|#1\right|}
\newcommand{\Tr}{{\text{Tr}}}
\newcommand{\tr}{{\text{\,tr}}}
\newcommand{\sym}{${\cal N}=4$ SYM  }

\preprint{  NORDITA 2019-056  \;  UUITP -- 21/19   }

\title{Integrable Fishnet from $\gamma$-Deformed $\mathcal N=2$ Quivers}

\author{Antonio Pittelli$^{a}$ and Michelangelo Preti$^{b}$}

\affiliation{%
\\
\(^{a}\) Department of Physics and Astronomy, Uppsala University, Box 516, SE-75120 Uppsala, Sweden
 \\
\(^{b}\) Nordita, KTH Royal Institute of Technology and Stockholm University, Roslagstullsbacken 23, SE-10691 Stockholm, Sweden
\\
}

\begin{abstract}
We introduce  bi-fermion fishnet theories, a class of models  describing integrable sectors of four-dimensional gauge theories with non-maximal supersymmetry. Bi-fermion theories are characterized by a single complex scalar field and two Weyl fermions interacting only via chiral Yukawa couplings. The latter generate oriented Feynman diagrams  forming  hexagonal lattices, whose fishnet structure signals an underlying integrability that we exploit to compute anomalous dimensions of BMN-vacuum operators.  Furthermore, we investigate Lunin-Maldacena deformations of $\mathcal N=2$ superconformal field theories with deformation parameter $\gamma$ and prove that  bi-fermion models emerge  in the limit of large imaginary $\gamma$ and vanishing ’t Hooft coupling $g$, with $g \, e^{-\f\im2 \gamma}$ fixed. Finally, we explicitly find non-trivial conformal fixed points  and compute the   scaling dimensions of operators for any $\gamma$ and in presence of double-trace deformations.
\end{abstract}  

 \maketitle

\section{Introduction and Conclusions}

Revealing  integrable sectors of a given quantum field theory (QFT) is an extremely compelling problem. Indeed, integrability allows for the exact computation of non-trivial physical observables and uncovers elegant mathematical structures as well as surprising correspondences. Established instances of integrability in QFT are the duality between the exactly solvable Heisenberg spin chain and QCD in the high-energy limit  \cite{Lipatov:1993yb, Faddeev:1994zg, Korchemsky:1994um}, the Bethe Ansatz for  $\mathcal N=4$ SYM \cite{Minahan:2002ve} and  for ABJM  \cite{Aharony:2008ug, Minahan:2008hf}.   Recently, a novel integrable 4d QFT was proposed \cite{Gurdogan:2015csr}. This theory, called $\chi{\rm CFT}_4$, was obtained by  considering the double-scaling (DS) limit \footnote{Similar double scaling limits were also studied in \cite{Correa:2012nk,Bonini:2016fnc,deLeeuw:2016vgp,Aguilera-Damia:2016bqv,Preti:2017fhw,Cavaglia:2018lxi} in different contexts.} of  $\gamma$-deformed $\mathcal N=4$ SYM, namely the regime of strong imaginary twists $\gamma$ and weak 't Hooft coupling $g={g_{\rm YM}\sqrt{N_c}}/{4\pi}$. Although  $\chi{\rm CFT}_4$ is non-supersymmetric, non-unitary and has no gauge symmetry, it exhibits a non-trivial infrared fixed point and is integrable thanks to the fishnet structure of its Feynman diagrams \cite{Zamolodchikov:1980mb, Kazakov:2018gcy}.  Similar $\gamma$-deformations and DS limits were introduced for ABJM theory  \cite{Caetano:2016ydc}.

Both $\chi{\rm CFT}_4$ and its 3d companion descend from integrable QFTs,  $\mathcal N=4$ SYM and ABJM respectively. In this letter we make a step forward, by showing that the DS limit generates an integrable QFT even out of a \emph{non}-integrable one. We shall focus on 4d  $\mathcal N=2$ superconformal field theories (SCFT) with Lagrangian description, which in principle are not integrable.  

The archetypal  4d $\mathcal N=2$ SCFT is  superconformal QCD (SCQCD). It  consists of a vector multiplet in the adjoint representation of the gauge group $SU(N_c)$ and  $N_f = 2 \, N_c$ fundamental hypermultiplets. SCQCD is quantum conformal and its  $\beta$-function remains vanishing in the planar Veneziano limit $N_c, N_f \to \infty$ with $N_f / N_c = 2$. Importantly, SCQCD is amenable to a Lunin-Maldacena $\gamma$-deformation \cite{Lunin:2005jy}, breaking its $U(1)_r \times SU(2)_R$ R-symmetry to the corresponding Cartan subalgebra.  The $\gamma$-deformed vector and hypermultiplet Lagrangians are reported in (\ref{gamma-defVM}) and (\ref{gamma-defHM}), respectively.  Especially,  in the DS limit \cite{Gurdogan:2015csr} 
  \begin{align}\label{eq: doublescalinglimit}
& \gamma\to\im\infty ~ , \qquad   g\to0 ~ ,  \qquad  \xi := e^{-\im\tfrac{\gamma}{2}} ~ g ={\rm fixed} ~ ,
\end{align}
both  hypermultiplets and gauge fields  decouple, giving the \emph{bi-fermion theory}
\begin{align}
\label{eq: dsvmlag}
 \mathcal L^\xi_{\phi\lambda} \!\!=\!N_c{\rm Tr}\!\left[
  \im  \overline\lambda_{I } \slashed{\partial} {\lambda^I }\!\!-\!\frac{|\partial  \phi|^2}{2} \!+\! 4\pi\im\, \xi \!\p{ \lambda^1 \overline \phi \lambda^2 \!+\! \overline \lambda_1 \phi \overline \lambda_2  } \!\right],
\end{align} 
where the complex scalar $\phi$ and the  Weyl fermions $\lambda^I_\alpha,\overline \lambda_I^{\dot\alpha}$ transform in the adjoint representation of $SU(N_c)$. Alike $\chi{\rm CFT}_4$,  (\ref{eq: dsvmlag}) generates oriented (chiral)  fishnet Feynman diagrams,  which prevent self-energy corrections at any order of the coupling.  The underlying integrability of the subsector $(\phi, \lambda)$  correctly matches \cite{Gadde:2010ku, Pomoni:2011jj, Liendo:2011xb, Liendo:2011wc, Gadde:2012rv, Pomoni:2013poa}: there, a spin-chain picture was found, with $\lambda,\overline \lambda$ behaving as excitations over the BMN vacuum ${\rm Tr}\phi^L$, similarly to the case of $\mathcal N=4$ SYM. The RG stability of a model with only Yukawa couplings such as (\ref{eq: dsvmlag}) was studied in  \cite{Mamroud:2017uyz}.

\begin{figure}[!t]
\includegraphics[trim={0.5cm 8cm 2cm 8cm},width=5.5cm]{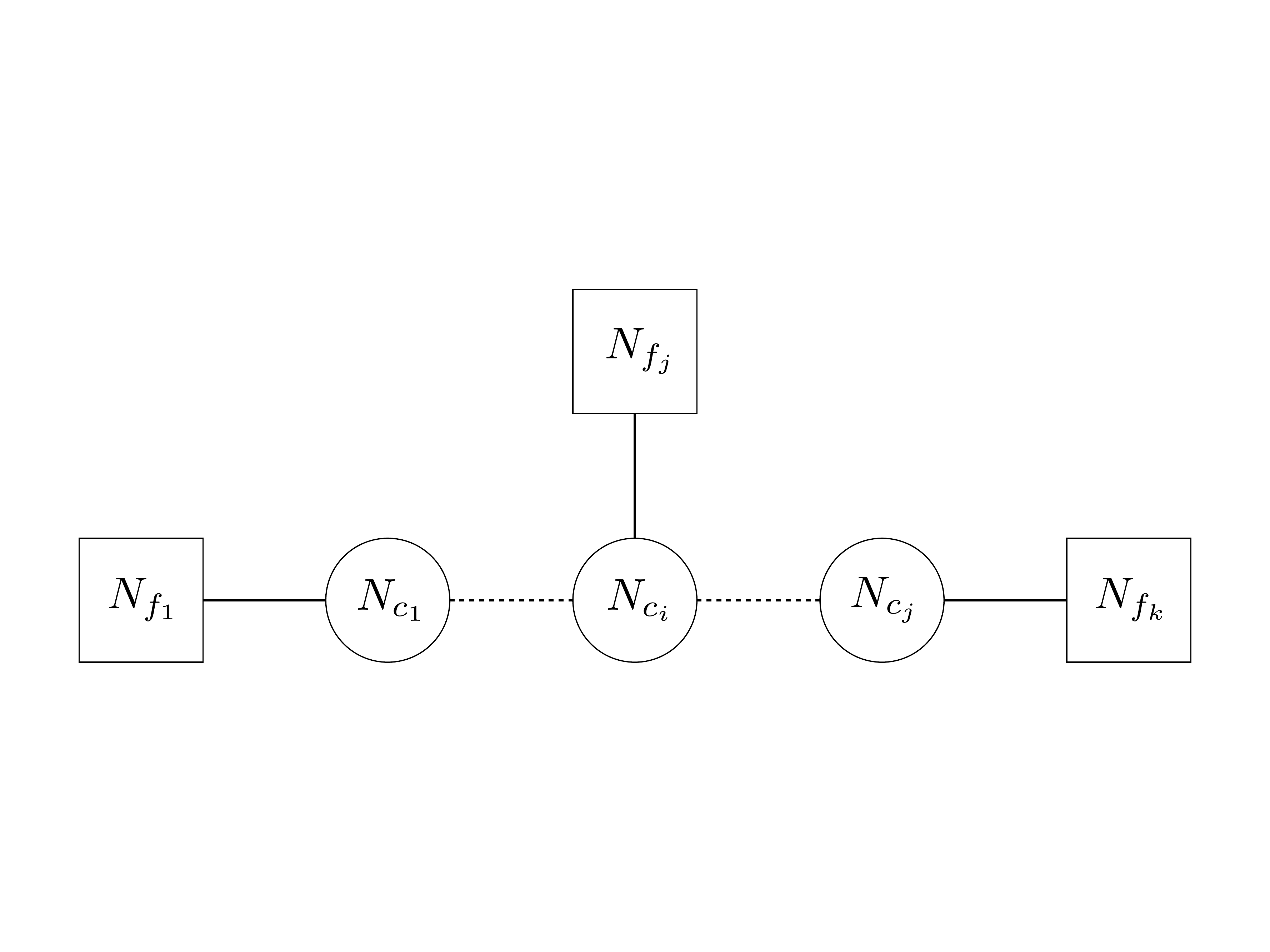}
 \caption{ A linear quiver.  Circles with label $N_{c_i}$ represent $SU(N_{c_i})$ vector multiplets. Lines connecting a $N_{c_i}$ circle to a $N_{f_j}$ square are $N_{f_j}$  hypers in the fundamental of  $SU(N_{c_i})$. Lines between  $N_{c_i}$ and $N_{c_j}$ circles are $SU(N_{c_i}) \times SU(N_{c_j})$ bifundamental hypers.  }
  \label{fig:quiver}
 \end{figure}

 $\mathcal N=2$ SCFTs with Lagrangian description are classified by quiver diagrams \footnote{For a complete classification of $\mathcal N=2$ SCFTs, please see  \cite{Bhardwaj:2013qia} and references therein.}, see Figure \ref{fig:quiver}.   A quiver is conformal if its vector multiplets have vanishing $\beta$-functions,  namely if a $N_i$-circle  is linked to a collection of objects with labels $M_i$ such that $\sum_i M_i = 2 N_i$. Simple examples are linear and circular quivers with all  circles having the same label. Up to double trace deformations, the $\gamma$-deformation commutes with the quiver structure: just like the undeformed one, the $\gamma$-deformed quiver is built by  suitably combining  (\ref{gamma-defVM}) and (\ref{gamma-defHM}).  However, the action of the DS limit on a $\gamma$-deformed  quiver is dramatic: matter multiplets decouple, leaving a collection of disjoint nodes captured by   bi-fermion theories with couplings $\xi_i$. Consequently, the DS limit  singles out the integrable sector of the quiver.
 
Several intriguing directions are yet to be explored. One is the study of  four-point functions of twist-two bosonic operators in the same spirit of \cite{Grabner:2017pgm,Gromov:2018hut,Kazakov:2018gcy}. The simple iterative form of the perturbative expansion can be encoded in a Bethe-Salpeter resummation in terms of  eigenvalues of  Hamiltonian evolution operators. Due to the underlying conformal symmetry, such eigenvalues can be computed via star-triangle relations \cite{Preti:2018vog,Preti:2019rcq}.  Another is the $D$-dimensional generalization of bi-fermion theories, as in \cite{Kazakov:2018qez}. 

Since the 4d  $\mathcal N = 2 $ and the 3d $\mathcal N = 4 $ superconformal algebras are related by dimensional reduction, it is natural to wonder about gamma-deformations and DS limit in one-dimension lower.
The $\gamma$-deformation acting on 3d $\mathcal N = 4 $ SCFTs breaks the R-symmetry $SU(2)_C \times SU(2)_H$ to the Cartan subalgebra $U(1)_C \times U(1)_H$. In turn,  the DS limit (\ref{eq: doublescalinglimit})   decouples gauge fields and matter multiplets. 
The DS limit of 3d SYM has a Yukawa interaction, while that of Chern-Simons matter theories produces quartic interactions of the type examined in \cite{Mamroud:2017uyz}. Generalizing to a quiver is straightforwardly achieved by considering multiple copies of the  doubly-scaled 3d Lagrangian. 
We shall present those results in the near future \cite{Pittelli:20191}.  

Another  is the holographic dual, which  would grant access to the strong-coupling regime via AdS/CFT. Gravity duals of $\mathcal N=2$ SCFT were constructed  \cite{Gaiotto:2009gz} and the corresponding $\gamma$-deformations recently appeared in \cite{Nunez:2019gbg}.  However, it is not clear how to perform the DS limit on the $\gamma$-deformed supergravity background without trivializing  the  gravity description. Since $\mathcal L^\xi_{\lambda\phi}$ is a fishnet theory, a more direct path to  holography would  be including fermionic interactions in the sigma-model of \cite{Basso:2018agi} or in the holographic fishchain of \cite{Gromov:2019aku}. Also, integrability on the QFT side should  imply some form of dual superconformal symmetry on the gravity side \cite{Berkovits:2008ic,Beisert:2008iq,Abbott:2015mla,Colgain:2016gdj}.  

Furthermore, it would be  fascinating to apply  the  Quantum spectral curve (QSC) method to the fishnet theory introduced in this letter. Although  there exists no QSC for $\mathcal N=2 $ SCFT, one may try to adapt the QSC of $\mathcal N=4$ SYM \cite{Kazakov:2015efa,Gromov:2017cja,Kazakov:2018hrh} to $\mathcal L^\xi_{\lambda\phi}$. Besides, fishnet theories admit a quantum R-matrix, whose quantum group  and secret symmetries \cite{Matsumoto:2007rh, Pittelli:2014ria} should be accessible via RTT expansion \cite{Beisert:2014hya,Hoare:2015kla,Pittelli:2017spf} .

%%%%%%%%%%%%%%%%%%%%
%%%%%%%%%%%%%%%%%%%%
%%%%%%%%%%%%%%%%%%%%

\section{ $\gamma$-Deformation and Double-Scaling Limit of $\mathcal N=2$ SCFT}
 
   \begin{table}[!t]
 \begin{center}
\begin{tabular}{c | c c c c c c c c c c c c }
 &$A_\mu$ & $\phi$  &  $\lambda^1$ & $\lambda^2$   & $q^1$ & $q^2$ & $\psi$   \\ [0.5ex] 
\hline
$U(1)_r$ & $0$& $+2$    & $+1$ &  $+1$ & 0 & 0 & $-1$    \\
$U(1)_R$   & 0 & 0  &  $+\f12$ &  $-\f12$   & $+\f12$ & $-\f12$ & 0   
\end{tabular}
 \end{center}
 \caption{R-charge assignments.    }
\label{table: vmrc}
\end{table}

A dynamical $\mathcal N=2$  vector multiplet $(\phi,\overline \phi, A_\mu,\lambda^I,\overline  \lambda_I )$   contains a complex scalar $\phi$, a  gauge field $A_\mu$ and four Weyl fermions $\lambda^I,\overline \lambda_I$ of opposite chirality \footnote{Nowhere in this letter  a bar denotes complex conjugation.}, all in the adjoint representation of the gauge group $SU(N_c)$. The corresponding Lagrangian has Yukawa and quartic scalar interactions. We shall perform a $\gamma$-deformation of the vector multiplet Lagrangian  by replacing the ordinary product with the following  associative, non-commutative $*$-product \cite{Lunin:2005jy}:
\begin{equation}\label{eq: luninmaldacenaproduct}
f * g := e^{\im\f\gamma2\p{Q^1\comm f Q^2 \comm g - Q^2\comm f Q^1 \comm g}}fg\,,
\end{equation}
where $Q^i\comm\Phi$ is the $U(1)_i$ charge of the field $\Phi$.  $\mathcal N=2$ SCFTs have $U(1)_r \times SU(2)_R$ global symmetry. If we choose the $\gamma$-deformation of (\ref{eq: luninmaldacenaproduct}) to act upon $U(1)_r\times U(1)_R$, with   $U(1)_R$ being the Cartan subalgebra of $SU(2)_R $,  the R-symmetry is broken according to  $U(1)_r \times SU(2)_R  \to U(1)_r\times U(1)_R$ . This procedure yields the $\gamma$-deformed vector multiplet Lagrangian:
\begin{align}
\label{gamma-defVM}
&\mathcal L^\gamma_{\rm VM} \! =\!N_c {\rm Tr}\!\left[\im  \overline\lambda_I\slashed{\mathcal D} \lambda^I \!- \frac{F^2}4  \!-\f12 |\mathcal D  \phi |^2   \!+\! 4\pi^2g^2 \! \comm{\phi,\overline \phi}^2 \right.  \\
& \!\!\left. + 4\pi\im  g \!\comm{   e^{- \im\f\gamma 2} \!\left(   \lambda^1\overline \phi \,  \lambda^2 + \overline \lambda_1\phi \overline \lambda_2  \right)
\!- e^{ \im\f\gamma 2 }\!\left(  \overline \lambda_2  \phi \, \overline \lambda_1 +\lambda^2\overline \phi\lambda^1  \right)  } \right]\!.\nn
\end{align}
As in \cite{Gurdogan:2015csr}, we can now perform the corresponding DS limit  (\ref{eq: doublescalinglimit}), transforming the $\gamma$-deformed action (\ref{gamma-defVM}) into (\ref{eq: dsvmlag}). These actions do not preserve  supersymmetry.

 A dynamical $\mathcal N=2$  hypermultiplet  $(q^I,\psi,\overline \psi)$    contains an $SU(2)_R$ doublet of scalars $q^I$ and two  Weyl fermions $\psi,\overline \psi $ of opposite chirality, all in a representation $\mathcal R$ of the gauge group. From the R-charges in Table \ref{table: vmrc}, one finds that (\ref{eq: luninmaldacenaproduct}) only affects gluino Yukawa couplings:
\begin{align}\label{gamma-defHM}
\mathcal L_{\rm HM}^\gamma  = \p{\mathcal L_{\rm HM}}_{\lambda=\overline \lambda=0}& + 4\pi\im g\,N_c {\rm Tr} \left[e^{- \im\f\gamma 4}( \psi{\lambda}^{1}q_{1}-\overline \psi\,{ \overline \lambda}_1q^{1})     \right.\nn\\
&  \quad\left.+ e^{\im\f\gamma 4}(\psi {\lambda}^{2}q_{2}- \overline \psi\, { \overline \lambda}_2  q^{2})\right]. 
\end{align}
In the DS limit (\ref{eq: doublescalinglimit}),  $\mathcal L_{\rm HM}^\gamma$ becomes the Lagrangian of a free theory  and  decouples from the vector multiplet.
 
%%%%%%%%%%%%%%%%%%%%
%%%%%%%%%%%%%%%%%%%%
%%%%%%%%%%%%%%%%%%%%

\section{Conformal symmetry}

In the quiver, every vector multiplet in the adjoint of $SU(N_{c})$ needs a double trace term 
\begin{equation}\label{Ldt}
\mathcal L_{dt} = (4\pi \alpha)^2\,{\rm Tr}[\phi^2]{\rm Tr}[\bar{\phi}^2]\,,
\end{equation}
for conformal symmetry to hold  quantum-mechanically. Here, $\alpha$ and $\phi$ respectively are the new induced coupling and the scalar field of the considered vector multiplet. The $\gamma$-deformed $\mathcal{N}=2$ theories supplied with $\mathcal L_{dt} $ remain unitary as long as the twist $\gamma$ and the couplings $(g,\alpha)$ are real. On the other hand, these are not CFT anymore since the double-trace coupling runs with the scale developing a non-vanishing $\beta$-function.

We can obtain the one-loop $\beta$-function similarly to the $\mathcal{N}=4$ counterpart \cite{Fokken:2013aea}, writing the bare double-trace term \eqref{Ldt} in terms of the renormalized one, where the coupling and the fields  counterterms appear in perturbation theory in $D=4-2\epsilon$
\begin{equation}\label{bare}
\alpha_0^2\,{\rm Tr}[\phi_0^2]{\rm Tr}[\bar{\phi}_0^2]=
\mu^{2\epsilon} (\alpha^2+\delta\alpha^2)\,{\rm Tr}[\phi^2]{\rm Tr}[\bar{\phi}^2]\,,
\end{equation}
where $\alpha^2=O(g^2)$ in order to cancel the UV divergences generated by the single-trace terms. We introduced the mass $\mu$ that rescales the unrenormalized 't Hooft coupling $g_0=\mu^\epsilon g$. The renormalized coupling and fields are defined as follows
\begin{equation}\label{Z}
\alpha^2=\mathcal{Z}_{\alpha^2}^{-1}\alpha_0^2\,,
\qquad\phi=\mathcal{Z}_{\phi}^{-1/2}\phi_0\,,
\end{equation}
with $\mathcal{Z}_{\alpha^2}=1+\delta_{\alpha^2}$ and $\mathcal{Z}_{\phi}=1+\delta_{\phi}$. Inserting these expressions in \eqref{bare} and using the independence of the bare couplings from $\mu$, at leading order we obtain
\begin{equation}\label{betadef}
\beta_{\alpha^2}\equiv \mu \frac{d}{d\mu}\alpha^2=2\epsilon(\delta\alpha^2-2\delta_{\phi} \alpha^2)\,.
\end{equation}

%\begin{figure}[!t]
%     \subfigure[]
%   {\includegraphics[width=1cm]{beta1}}
%   \hspace{.6cm}
%     \subfigure[]
%   {\includegraphics[width=1cm]{beta2}}
%      \hspace{.6cm}
%     \subfigure[]
%   {\includegraphics[width=1cm]{beta3}}
%      \hspace{.6cm}
%     \subfigure[]
%   {\includegraphics[width=1cm]{beta4}}
%      \hspace{.6cm}
%     \subfigure[]
%   {\includegraphics[width=1cm]{beta5}}
% \caption{Diagrams contributing to the renormalization of the double-trace coupling. Thick lines represent scalar propagators, dashed lines fermionic propagators and wavy lines gauge one. Black dots stand for double-trace vertices \eqref{Ldt}.}
%  \label{fig:betadiag}
% \end{figure}

\begin{figure}[!t]
\includegraphics[width=0.47\textwidth]{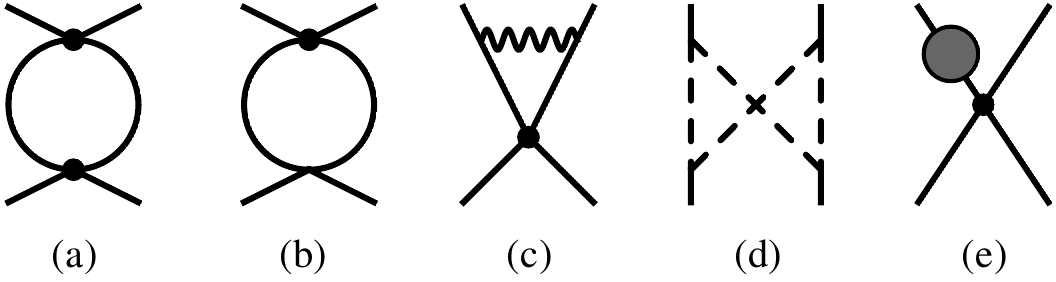}
 \caption{Diagrams contributing to the renormalization of the double-trace coupling. Thick lines represent scalar propagators, dashed lines fermionic propagators and wavy lines gauge one. Black dots stand for double-trace vertices \eqref{Ldt}.}
  \label{fig:betadiag}
 \end{figure}

Since in $\mathcal{N}=2$ SCFT all divergent contributions to the double-trace coupling $\alpha^2$ vanish, one can consider only the diagrams sensitive to the $\gamma$-deformation and hence deviate from their un-deformed counterparts. At one-loop, the only deformed terms in the action \eqref{gamma-defVM} contributing to the renormalization of $\alpha^2$ are the cubic Yukawa fermion-scalar couplings. The deformed vertices of the hypermultiplet action (\ref{gamma-defHM}) start  contributing at higher-order in perturbation theory.  The only diagrams at large $N_c$ relevant to the interaction \eqref{Ldt} are listed in Figure  \ref{fig:betadiag}. They depend on a single scalar one-loop integral $I_1$ given by
\begin{equation}
I_1\equiv\vcenter{\hbox{\includegraphics[width=1.5cm]{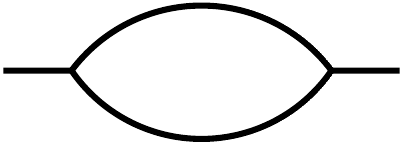}}}=\frac{\mu^{4-D}}{p^{2(2-D/2)}}G(1,1)\,,
\end{equation}
where the $G$-function is defined as follows \cite{Chetyrkin:1980pr,Vladimirov:1979zm}
\begin{equation}\label{Gfunction}
G(a,b)=\frac{\Gamma(a+b-D/2)\Gamma(D/2-a)\Gamma(D/2-b)}{\Gamma(a)\Gamma(b)\Gamma(D-a-b)}\,.
\end{equation}

The UV divergence of the one-loop renormalized double-trace vertex is given by the sum of the single UV divergencies of the diagrams (a), (b), (c) and (d) of FIG \ref{fig:betadiag}. They appear in different fashions reflecting them respect to the vertical or horizontal axes, then the total contribution is given by
\begin{equation}\label{UV1}
\delta\alpha^2\!=\!K\!\left[\vcenter{\hbox{\includegraphics[width=.45cm]{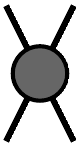}}}\right]\!=\!2(\alpha^4\!-\!g^2\alpha^2(1\!+\!a)\!+\!2g^4\sin^2\gamma)K\!\left[I_1\right],
\end{equation}
where the operator $K$ extracts the UV divergence that it appears as a pole in $\epsilon$ and $a$ is an unspecified gauge-fixing parameter. In \eqref{UV1} we considered also the divergent contribution given by the deformation-independent diagrams, corresponding to setting $\alpha$ and $\gamma$ to zero in the deformation-dependent terms.

The divergent contribution coming from the wave function renormalization is given by the diagram (e) of Figure \ref{fig:betadiag}. This diagram appears in four different fashions depending on the position of the one-loop corrected bosonic propagator. The bosonic self-energy contains corrections given by two gauge-boson or Yukawa three-vertices. In momentum space, the divergence is factorized then we obtain
\begin{equation}\label{UV2}
\delta_{\phi}\alpha^2=K\!\left[\vcenter{\hbox{\includegraphics[width=.8cm]{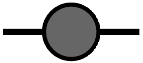}}}\right]\frac{1}{p^2}\vcenter{\hbox{\includegraphics[width=.6cm]{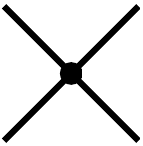}}}=-2g^2\alpha^2(1+a)K\!\left[I_1\right]\,.
\end{equation}
Finally, summing up the two contributions \eqref{UV1} and \eqref{UV2} as in \eqref{betadef}, we have
\begin{equation}\label{betagamma}
\beta_{\alpha^2}= 4(2g^4 \sin^2\gamma +\alpha^4)+\dots\,,
\end{equation}
where the dependence on the gauge-fixing parameter $a$ has canceled as expected. For generic value of $\alpha^2$, the one-loop coupling renormalization leads to a non-vanishing $\beta$-function, and conformal invariance is broken. However, we can impose the vanishing of \eqref{betagamma} obtaining the following complex conjugate fixed points
\begin{equation}\label{fixedpoint}
\alpha_{\pm}^2=\pm\, i \sqrt{2}\,g^2\,\sin\gamma+O(g^4)\,.
\end{equation}

The $\beta$-function \eqref{betagamma} is quadratic as predicted in \cite{Dymarsky:2005uh,Pomoni:2008de,Pomoni:2010et} for any large-$N$ CFT with a non-running single-trace coupling and a running double-trace one. In general, 
\begin{equation}\label{betagen}
\beta_{\alpha^2}=a(g)\alpha^4+b(g)\alpha^2+c(g)\,,
\end{equation}
where $a$, $b$ and $c$ are functions of the coupling. At the fixed points $\alpha_{\pm}^2$, the anomalous dimension of the twist-two operator $\mathcal{O}_2=\text{Tr}[\phi^2]$ can be expressed in terms of the discriminant of $\beta_{\alpha^2}=0$ such that
\begin{equation}\label{discriminant}
4\gamma^2_{\mathcal{O}_2}=b^2-4ac\quad\Rightarrow\quad
\gamma_{\mathcal{O}_2}=\mp 4\,i \sqrt{2}\,g^2\,\sin\gamma\,.
\end{equation}

Performing the DS limit \eqref{eq: doublescalinglimit}, one can obtain from \eqref{betagamma} the $\beta$-function for the bi-fermion theory
\begin{equation}
\beta_{\alpha^2}^{\rm DS}= 4\alpha^4-2\xi^4+\dots\,,
\end{equation}
and from \eqref{fixedpoint} and \eqref{discriminant}, the fixed-points and anomalous dimension of the length-two operator $\mathcal{O}_2$
\begin{equation}\label{DSprediction}
\alpha_{\pm}^{\rm DS}\!=\!\pm\frac{\xi^2}{\sqrt{2}}+O(\xi^4)\quad\!\text{and}\!\quad\gamma_{\mathcal{O}_2}^{\rm DS}\!=\!\mp 2\sqrt{2}\xi^2+O(\xi^6)\,.
\end{equation}

 %%%%%%%%%%%%%%%%%%%%
%%%%%%%%%%%%%%%%%%%%
%%%%%%%%%%%%%%%%%%%%

\section{Wheel diagrams}

\begin{figure}[!t]
\subfigure[]
{\includegraphics[trim={6cm 9cm 6cm 6cm},clip,width=3.5cm]{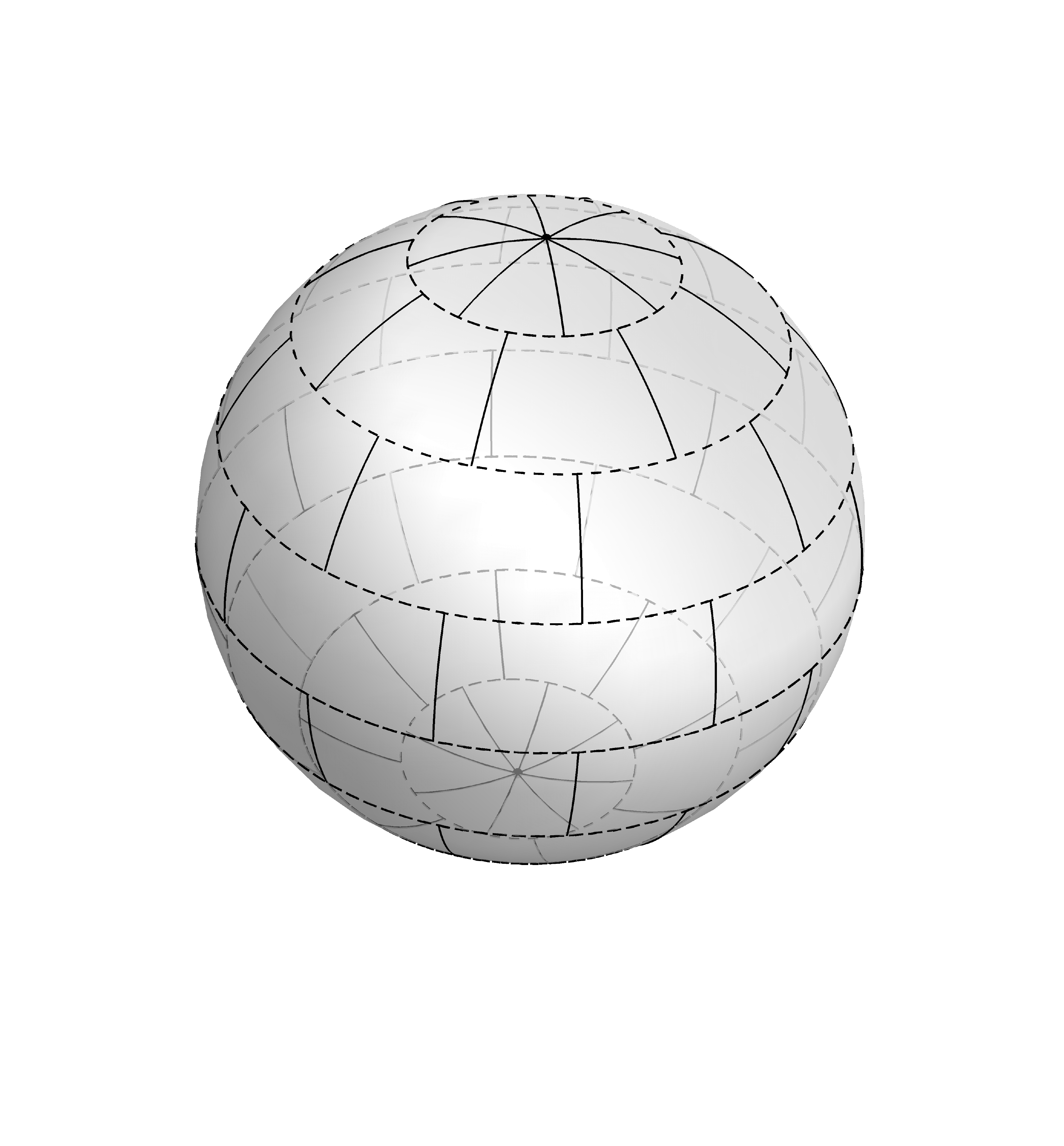}}
\hspace{.2cm}
     \subfigure[]
   {\includegraphics[width=3.4cm]{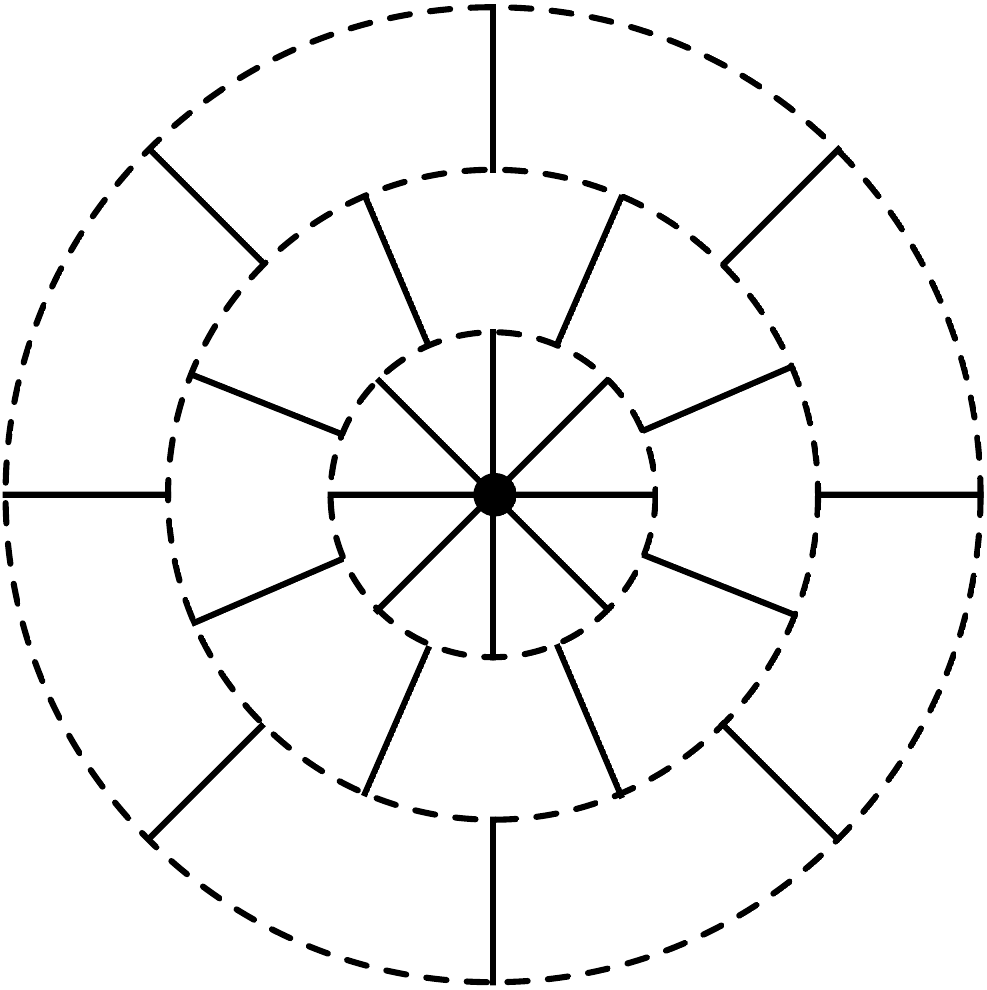}}
\caption{The only diagram generated by (\ref{eq: dsvmlag})  for  the two-point function of  BMN vacuum operators sitting at the poles of the sphere in (a).  Dashed lines are gluino propagators while   solid lines are scalar ones.  The divergence of this graph is encoded in the wheel graph depicted  in (b). }
\label{fig:sphere}
\end{figure}

The simplest single-trace bosonic operator of length-$L$ in a $\gamma$-deformed quiver theory is
\begin{equation}\label{OL}
\mathcal{O}_L\equiv\text{Tr}[\phi^L]\,,
\end{equation}
where $L\geq 2$ for $SU(N_c)$. This operator is protected in the corresponding undeformed $\mathcal{N}=2$ SYM theory where its scaling dimension is uniquely determined by its global charge $L$.  
The $\gamma$-deformation produces a non-trivial anomalous dimension to its scaling dimension. In particular at one-loop it is generated by wrapping diagrams with a single matter-type wrapping loop and, for $L=2$ case, by diagrams in which the double-trace coupling \eqref{Ldt} occurs. In both cases vertices coming from the hypermultiplet lagrangian \eqref{gamma-defHM} do not contribute.

The composite operator \eqref{OL} needs to be renormalized. Then, following \cite{Fokken:2014soa}, we introduce a renormalization constant that expresses the renormalized operator in terms of the bare one 
\begin{equation}
\mathcal{O}_L(\phi)=\mathcal{Z}_{\mathcal{O}_L}\mathcal{O}_{L,0}(\phi_0)\,.
\end{equation}
This definition, together with \eqref{Z} and the renormalization group equations, leads to the following anomalous dimension
\begin{equation}\label{gammaL}
\gamma_{\mathcal{O}_L}\!\equiv\! -\mu \frac{\partial}{\partial \mu}\log\mathcal{Z}_{\mathcal{O}_L}\!\!=\!\left(\!\epsilon g\frac{\partial}{\partial g}-\beta_{\alpha^2}\frac{\partial}{\partial \alpha^2}\!\right)\!\log\mathcal{Z}_{\mathcal{O}_L}.
\end{equation}
Since the above result must be finite when $\epsilon\rightarrow 0$, at the fixed points \eqref{fixedpoint} the function $\log\mathcal{Z}_{\mathcal{O}_L}$ cannot contain poles higher than $1/\epsilon$.

In the $L\geq 3$ case, the double-trace vertex \eqref{Ldt} doesn't contribute, then the only deformation-dependent diagram at the lowest order is the one in Figure \ref{fig:sphere}(b) with one fermionic wrapping. This diagram can be written in terms of the scalar integral $P_L$ with the following UV divergence \cite{Broadhurst:1985vq}
\begin{equation}
K[P_L]\!\equiv\! K\!\!\left[\vcenter{\hbox{\includegraphics[width=1.2cm]{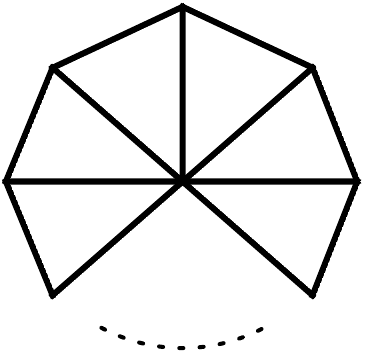}}}\right]\!\!=\!\frac{1}{\epsilon}\frac{2}{L}\binom{2L-3}{L-1}\zeta_{2L-3}\,,
\end{equation}
where $\zeta_{n}$ is the Riemann zeta function $\zeta(n)$. 
The renormalization constant $\mathcal{Z}_{\mathcal{O}_L}$ is given by the negative sum of all the poles, then considering that the fermionic wrapping appears with two opposite phases from the Lagrangian \eqref{gamma-defVM}, we have
\begin{equation}\label{ZL}
\mathcal{Z}_{\mathcal{O}_L}\equiv1+\delta\mathcal{Z}_{\mathcal{O}_L}
=1-8 g^{2L}\sin^2L\frac{\gamma}{2}K[P_L]\,,
\end{equation}
where we considered also the divergence of the undeformed diagrams simply subtracting the deformation dependent $\delta\mathcal{Z}_{\mathcal{O}_L}$ computed for $\gamma=0$. Then using \eqref{gammaL}, for $L\geq 3$ we obtain
\begin{equation}
\gamma_{\mathcal{O}_L}=- 32\, g^{2L}\sin^2{L\frac{\gamma}{2}}\,\binom{2L-3}{L-1}\zeta_{2L-3}+\dots\,,
\end{equation}
and for the theory \eqref{eq: doublescalinglimit}
\begin{equation}
\gamma^{\rm DS}_{\mathcal{O}_L}=8\,\xi^{2L}\,\binom{2L-3}{L-1}\zeta_{2L-3}+\dots\,,
\end{equation}
where we used the DS limit \eqref{eq: dsvmlag}.

The $L=2$ case is more involved. Indeed one has to consider contributions from the double-trace vertex in addition to the fermionic wrapping.
At one-loop the only contribution to the renormalization constant is given by a single double-trace vertex
\begin{equation}\label{Z1}
\delta\mathcal{Z}_{\mathcal{O}_2}^{\text{1-loop}}=2 \alpha^2 K[I_1]\,.
\end{equation}
At two-loop  the diagram containing only single-trace couplings is given by the fermionic wrapping considered above. Its contribution corresponds to set $L=2$ in \eqref{ZL} but substituting $P_2$, that contains an IR divergence, with the following integral
\begin{equation}
I_2\equiv\vcenter{\hbox{\includegraphics[width=1.5cm]{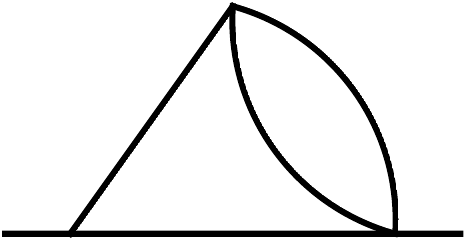}}}=\frac{\mu^{2(4-D)}}{p^{4(2-D/2)}}G(1,1)G(3-\tfrac{D}{2},1)\,,
\end{equation}
where $G$ is given by \eqref{Gfunction}. The diagrams depending on the double-trace couplings are given by the contraction of the operator $\mathcal{O}_2$ with the graph in Figure \ref{fig:betadiag}. Those diagrams together with the ones containing the one-loop counterterms associated to the renormalization of the operator, mass and double-trace coupling (given by \eqref{UV1}) lead to
\begin{equation}\label{Z2}
\delta\mathcal{Z}_{\mathcal{O}_2}^{\text{2-loop}}=4 \alpha^4 K[I_1]^2-8g^4\sin^2\gamma K[I_2-K[I_1]I_1].
\end{equation}
Summing up \eqref{Z1} and \eqref{Z2} and using \eqref{gammaL} considering that $\alpha^2=O(g^2)$, we obtain
\begin{equation}
\gamma_{\mathcal{O}_2}=4\,\alpha^2-16\, g^{4}\sin^2{\gamma}+\dots\;.
\end{equation}
It is easy to check that at the fixed points \eqref{fixedpoint}, this result gives the expected one-loop anomalous dimension \eqref{discriminant}.

The exact anomalous dimension of the operator $\mathcal{O}_2$ can be computed with the same technique of \cite{Grabner:2017pgm,Gromov:2018hut,Kazakov:2018gcy}. Diagonalizing the fermionic graph-building hamiltonian ( the same appearing in \cite{Kazakov:2018gcy}) and performing the Bethe-Salpeter resummation, we obtain \cite{Pittelli:20192}
\begin{equation}
\gamma_{\mathcal{O}_2}^{\rm DS}\!=\!\mp 2\sqrt{2}\!\left[\xi^2\!+\! (1\!+\!3\zeta_3)\xi^6\!+\! \tfrac{7+9\zeta_3(2+3\zeta_3)+30\zeta_5}{2}\xi^{10}\!+\!\dots\right]\!,
\end{equation}
matching our prediction \eqref{DSprediction}.

 %%%%%%%%%%%%%%%%%%%%
%%%%%%%%%%%%%%%%%%%%
%%%%%%%%%%%%%%%%%%%%

\section{Integrability and Fishnet}

It was argued that any $\mathcal N=2$ SCFTs  has an integrable subsector  in the planar Veneziano limit \cite{Gadde:2010ku, Pomoni:2011jj, Liendo:2011xb, Liendo:2011wc, Gadde:2012rv, Pomoni:2013poa}. This sector has a spin-chain interpretation in terms of a vacuum $\mathcal O_{L} = {\rm Tr}\phi^L$ and elementary  excitations (magnons) corresponding to insertions of gluinos $\lambda$ and of gauge-covariant derivatives $\mathcal D_{\alpha\dot\alpha}$. Gauge fields decouple in the DS limit, leaving  gluinos as the only  non-trivial excitations. In  undeformed SCFTs,   the vacuum $ \mathcal  O_{L}  $ has zero anomalous dimension as it  satisfies the BPS condition $\Delta =|r|/2$, with $r = 2 L$ being the $U(1)_r$  charge. The one-magnon excitation is encoded in the operator $\mathcal O_\lambda =  {\rm Tr}\comm{\lambda\phi^{L-1}}$, with bare dimension $\Delta^{(0)}_{\mathcal O_\lambda } =\p{3/2}+\p{ L-1}$ and $U(1)_r$ charge $r =1+2\p{ L-1}$ satisfying $\Delta^{(0)}_{\mathcal O_\lambda }  - \p{r/2} = 1$. Up to the  coupling constant redefinition  $g^2\to f(g)=g^2 + O(g^3)$, integrable excitations  have  dispersion relations  analogous to their $\mathcal N=4 $ SYM counterparts \cite{Pomoni:2013poa}. By using $p\sim \gamma$  \cite{Caetano:2016ydc},  the DS limit   removes $O(g^3)$ corrections:
\begin{align}
\gamma_{\mathcal O_\lambda }  \! = -1 \!+\! \sqrt{ 1 \!+\! 16   f(g) \sin^2\f p2} \to -1\!+\!  \sqrt{ 1 -4 \, \xi^2 } ~ .  
\end{align} 
 
The integrability of gluino excitations above the BMN-like vacuum ${\rm Tr}\phi^L$ can be detected without resorting to the original undeformed theory.  Indeed, for BMN-like external state, the bi-fermion  Feynman graphs are brick-wall  fishnet diagrams, see Figure \ref{fig:sphere}. These are known  to enjoy features characterizing integrable models, such as Yangian invariance \cite{Chicherin:2017frs} and star-triangle relations  \cite{Zamolodchikov:1980mb}.  Technically, the Hamiltonian of the dual integrable system is related to  the graph building operator acting on the planar wheel diagram.   
The Hamiltonian  and the  transfer matrix  for a theory with brick-wall domain graphs were explicitly showed to be commuting \cite{Kazakov:2018gcy}, which demonstrates the integrability of the system. In fact,  the Hamiltonian is one among an infinite tower of conserved charges encoded in the transfer matrix. 

%%%%%%%%%%%%%%%%%%%%
%%%%%%%%%%%%%%%%%%%%
%%%%%%%%%%%%%%%%%%%%

\begin{acknowledgments}
\section*{Acknowledgments}
\label{sec:acknowledgments}
We would very much like to thank N. Gromov, V. Kazakov, E. Pomoni  and K. Zarembo for illuminating discussions and critical reading of the manuscript. AP is supported by the ERC STG Grant 639220.
\end{acknowledgments}

%%%%%%%%%%%%%%%%%%%%
%%%%%%%%%%%%%%%%%%%%
%%%%%%%%%%%%%%%%%%%%

\bibliographystyle{MyStyle}
\bibliography{Bifermion}

\begin{thebibliography}{10}
\providecommand{\url}[1]{\texttt{#1}}
\providecommand{\urlprefix}{URL }
\providecommand{\eprint}[2][]{\url{#2}}

\bibitem{Lipatov:1993yb}
L.~N. Lipatov, JETP Lett. \textbf{59}, 596, [Pisma Zh. Eksp. Teor.
  Fiz.59,571(1994)] (1994), \eprint{hep-th/9311037}.

\bibitem{Faddeev:1994zg}
L.~D. Faddeev and G.~P. Korchemsky, Phys. Lett. \textbf{B342}, 311 (1995),
  \eprint{hep-th/9404173}.

\bibitem{Korchemsky:1994um}
G.~P. Korchemsky, Nucl. Phys. \textbf{B443}, 255 (1995),
  \eprint{hep-ph/9501232}.

\bibitem{Minahan:2002ve}
J.~A. Minahan and K.~Zarembo, JHEP \textbf{03}, 013 (2003),
  \eprint{hep-th/0212208}.

\bibitem{Aharony:2008ug}
O.~Aharony, O.~Bergman, D.~L. Jafferis and J.~Maldacena, JHEP \textbf{10}, 091
  (2008), \eprint{0806.1218}.

\bibitem{Minahan:2008hf}
J.~A. Minahan and K.~Zarembo, JHEP \textbf{09}, 040 (2008), \eprint{0806.3951}.

\bibitem{Gurdogan:2015csr}
O.~G\"urdogan and V.~Kazakov, Phys. Rev. Lett. \textbf{117}, 201602 (2016),
  \eprint{arXiv:1512.06704 [hep-th]}.

\bibitem{Note1}
Similar double scaling limits were also studied in \cite
  {Correa:2012nk,Bonini:2016fnc,deLeeuw:2016vgp,Aguilera-Damia:2016bqv,Preti:2017fhw,Cavaglia:2018lxi}
  in different contexts.

\bibitem{Zamolodchikov:1980mb}
A.~B. Zamolodchikov, Phys. Lett. \textbf{97B}, 63 (1980).

\bibitem{Kazakov:2018gcy}
V.~Kazakov, E.~Olivucci and M.~Preti  (2018), \eprint{1901.00011}.

\bibitem{Caetano:2016ydc}
J.~Caetano, {\"O}.~G{\"u}rdo{\u{g}}an and V.~Kazakov, Journal of High Energy
  Physics \textbf{2018}, 77 (2018).

\bibitem{Lunin:2005jy}
O.~Lunin and J.~M. Maldacena, JHEP \textbf{0505}, 033 (2005),
  \eprint{arXiv:0502086 [hep-th]}.

\bibitem{Gadde:2010ku}
A.~Gadde and L.~Rastelli, JHEP \textbf{04}, 053 (2012), \eprint{1012.2097}.

\bibitem{Pomoni:2011jj}
E.~Pomoni and C.~Sieg  (2011), \eprint{1105.3487}.

\bibitem{Liendo:2011xb}
P.~Liendo, E.~Pomoni and L.~Rastelli, JHEP \textbf{07}, 003 (2012),
  \eprint{1105.3972}.

\bibitem{Liendo:2011wc}
P.~Liendo and L.~Rastelli, JHEP \textbf{10}, 117 (2012), \eprint{1111.5290}.

\bibitem{Gadde:2012rv}
A.~Gadde, P.~Liendo, L.~Rastelli and W.~Yan, JHEP \textbf{08}, 015 (2013),
  \eprint{1211.0271}.

\bibitem{Pomoni:2013poa}
E.~Pomoni, Nucl. Phys. \textbf{B893}, 21 (2015), \eprint{1310.5709}.

\bibitem{Mamroud:2017uyz}
O.~Mamroud and G.~Torrents, JHEP \textbf{2017} (2017), \eprint{arXiv:1703.04152
  [hep-th]}.

\bibitem{Note2}
For a complete classification of $\protect \mathcal N=2$ SCFTs, please see
  \cite {Bhardwaj:2013qia} and references therein.

\bibitem{Grabner:2017pgm}
D.~Grabner, N.~Gromov, V.~Kazakov and G.~Korchemsky  (2017),
  \eprint{arXiv:1711.04786 [hep-th]}.

\bibitem{Gromov:2018hut}
N.~Gromov, V.~Kazakov and G.~Korchemsky  (2018), \eprint{1808.02688}.

\bibitem{Preti:2018vog}
M.~Preti  (2018), \eprint{1811.04935}.

\bibitem{Preti:2019rcq}
M.~Preti, \textit{{}} (2019), \eprint{1905.07380}.

\bibitem{Kazakov:2018qez}
V.~Kazakov and E.~Olivucci, Phys. Rev. Lett. \textbf{121}, 131601 (2018),
  \eprint{1801.09844}.

\bibitem{Pittelli:20191}
A.~Pittelli and M.~Preti \eprint{to appear}.

\bibitem{Gaiotto:2009gz}
D.~Gaiotto and J.~Maldacena, JHEP \textbf{10}, 189 (2012), \eprint{0904.4466}.

\bibitem{Nunez:2019gbg}
C.~Núñez, D.~Roychowdhury, S.~Speziali and S.~Zacarías, Nucl. Phys.
  \textbf{B943}, 114617 (2019), \eprint{1901.02888}.

\bibitem{Basso:2018agi}
B.~Basso and D.-l. Zhong, JHEP \textbf{01}, 002 (2019), \eprint{1806.04105}.

\bibitem{Gromov:2019aku}
N.~Gromov and A.~Sever  (2019), \eprint{1903.10508}.

\bibitem{Berkovits:2008ic}
N.~Berkovits and J.~Maldacena, JHEP \textbf{09}, 062 (2008),
  \eprint{0807.3196}.

\bibitem{Beisert:2008iq}
N.~Beisert, R.~Ricci, A.~A. Tseytlin and M.~Wolf, Phys. Rev. \textbf{D78},
  126004 (2008), \eprint{0807.3228}.

\bibitem{Abbott:2015mla}
M.~C. Abbott, J.~Murugan, S.~Penati, A.~Pittelli, D.~Sorokin, P.~Sundin,
  J.~Tarrant, M.~Wolf and L.~Wulff, JHEP \textbf{12}, 104 (2015),
  \eprint{1509.07678}.

\bibitem{Colgain:2016gdj}
E.~O. Colgain and A.~Pittelli, Phys. Rev. \textbf{D94}, 106006 (2016),
  \eprint{1609.03254}.

\bibitem{Kazakov:2015efa}
V.~Kazakov, S.~Leurent and D.~Volin, JHEP \textbf{12}, 044 (2016),
  \eprint{arXiv:1510.02100 [hep-th]}.

\bibitem{Gromov:2017cja}
N.~Gromov, V.~Kazakov, G.~Korchemsky, S.~Negro and G.~Sizov, JHEP \textbf{01},
  095 (2018), \eprint{1706.04167}.

\bibitem{Kazakov:2018hrh}
V.~Kazakov 293--342, [Rev. Math. Phys.30,no.07,1840010(2018)] (2018),
  \eprint{1802.02160}.

\bibitem{Matsumoto:2007rh}
T.~Matsumoto, S.~Moriyama and A.~Torrielli, JHEP \textbf{09}, 099 (2007),
  \eprint{0708.1285}.

\bibitem{Pittelli:2014ria}
A.~Pittelli, A.~Torrielli and M.~Wolf, J. Phys. \textbf{A47}, 455402 (2014),
  \eprint{1406.2840}.

\bibitem{Beisert:2014hya}
N.~Beisert and M.~de~Leeuw, J. Phys. \textbf{A47}, 305201 (2014),
  \eprint{1401.7691}.

\bibitem{Hoare:2015kla}
B.~Hoare, A.~Pittelli and A.~Torrielli, Phys. Rev. \textbf{D93}, 066006 (2016),
  \eprint{1509.07587}.

\bibitem{Pittelli:2017spf}
A.~Pittelli, Nucl. Phys. \textbf{B935}, 271 (2018), \eprint{1711.02468}.

\bibitem{Note3}
Nowhere in this letter a bar denotes complex conjugation.

\bibitem{Fokken:2013aea}
J.~Fokken, C.~Sieg and M.~Wilhelm, J. Phys. \textbf{A47}, 455401 (2014),
  \eprint{arXiv:1308.4420 [hep-th]}.

\bibitem{Chetyrkin:1980pr}
K.~G. Chetyrkin, A.~L. Kataev and F.~V. Tkachov, Nucl. Phys. \textbf{B174}, 345
  (1980).

\bibitem{Vladimirov:1979zm}
A.~A. Vladimirov, Theor. Math. Phys. \textbf{43}, 417, [Teor. Mat.
  Fiz.43,210(1980)] (1980).

\bibitem{Dymarsky:2005uh}
A.~Dymarsky, I.~R. Klebanov and R.~Roiban, JHEP \textbf{08}, 011 (2005),
  \eprint{arXiv:0505099 [hep-th]}.

\bibitem{Pomoni:2008de}
E.~Pomoni and L.~Rastelli, JHEP \textbf{04}, 020 (2009), \eprint{0805.2261}.

\bibitem{Pomoni:2010et}
E.~Pomoni and L.~Rastelli, JHEP \textbf{10}, 171 (2012), \eprint{1002.0006}.

\bibitem{Fokken:2014soa}
J.~Fokken, C.~Sieg and M.~Wilhelm, JHEP \textbf{09}, 78 (2014),
  \eprint{arXiv:1405.6712 [hep-th]}.

\bibitem{Broadhurst:1985vq}
D.~J. Broadhurst, Phys. Lett. \textbf{B164}, 356 (1985).

\bibitem{Pittelli:20192}
A.~Pittelli and M.~Preti \eprint{to appear}.

\bibitem{Chicherin:2017frs}
D.~Chicherin, V.~Kazakov, F.~Loebbert, D.~Muller and D.-L. Zhong, Phys. Rev.
  \textbf{D96}, 121901 (2017), \eprint{1708.00007}.

\bibitem{Correa:2012nk}
D.~Correa, J.~Henn, J.~Maldacena and A.~Sever, JHEP \textbf{05}, 098 (2012),
  \eprint{1203.1019}.

\bibitem{Bonini:2016fnc}
M.~Bonini, L.~Griguolo, M.~Preti and D.~Seminara, JHEP \textbf{05}, 180 (2016),
  \eprint{1603.00541}.

\bibitem{deLeeuw:2016vgp}
M.~de~Leeuw, A.~C. Ipsen, C.~Kristjansen and M.~Wilhelm, Phys. Lett.
  \textbf{B768}, 192 (2017), \eprint{1608.04754}.

\bibitem{Aguilera-Damia:2016bqv}
J.~Aguilera-Damia, D.~H. Correa and V.~I. Giraldo-Rivera, JHEP \textbf{03}, 023
  (2017), \eprint{1612.07991}.

\bibitem{Preti:2017fhw}
M.~Preti, D.~Trancanelli and E.~Vescovi, JHEP \textbf{10}, 079 (2017),
  \eprint{1708.04884}.

\bibitem{Cavaglia:2018lxi}
A.~Cavagli\'{a}, N.~Gromov and F.~Levkovich-Maslyuk, JHEP \textbf{10}, 060
  (2018), \eprint{1802.04237}.

\bibitem{Bhardwaj:2013qia}
L.~Bhardwaj and Y.~Tachikawa, JHEP \textbf{12}, 100 (2013), \eprint{1309.5160}.

\end{thebibliography}

\end{document}